\documentstyle[amsfonts,amssymb,amsthm,amsmath,bbm,aps,times,rmp,prbbib,t1enc,twoside]{revtex}
\makeatletter
\newcommand{\nn}{n}
\renewcommand{\ll}{l} 
\let\Im\undefined

\DeclareMathOperator{\Im}{Im}

\DeclareMathOperator{\Tr}{Tr}
\def\indfkt#1{\raisebox{0.4ex}{$\chi$}_{#1}}
\newif\ifper\pertrue

\def\au#1#2{#1 #2}
\def\et{ and }
\def\bti#1{\emph{#1}}
\def\ti#1{``#1,''}

\def\z{\@ifnextchar[\zz\zzz}
\def\zz[#1]#2#3#4#5{\perfalse#2 \textbf{#3}, \linebreak[0]#4
  (#5) [#1].}
\def\zzz#1#2#3#4{#1 \textbf{#2}, \linebreak[0]#3 (#4)\ifper.\fi\pertrue}

\def\pub{\@ifstar\pubstar\pubnostar}
\def\pubnostar{\@ifnextchar[\@@pubnostar\@pubnostar}
\def\@@pubnostar[#1]#2#3#4{(#2, #3, #4), #1\ifper.\fi\pertrue}
\def\@pubnostar#1#2#3{(#1, #2, #3)\ifper.\fi\pertrue}
\def\pubstar[#1]#2#3#4{\perfalse (#2, #3, #4) [#1].}
\makeatother

\begin{document}
%
\title{Upper bounds on the density of states of single Landau levels
broadened by Gaussian random potentials}
%
%
\author{Thomas Hupfer, Hajo Leschke, and Simone Warzel}
\address{\footnotesize \it Institut f{\"u}r Theoretische Physik, 
                       Universit{\"a}t 
                       Erlangen-N{\"u}rnberg, Staudtstra{\ss}e 7,\\
                       D--91058 Erlangen, Germany\\[2ex]
                       {\rm October 2001.~ 
                        To appear in~ \emph{J. Math. Phys.}
                         }}
\maketitle
\begin{abstract}
\begin{minipage}[t]{10.3cm}
  {\small
We study a non-relativistic charged particle on the Euclidean plane $ {\mathbbm{R}}^2 $
subject to a perpendicular constant magnetic field and an $ {\mathbbm{R}}^2 $-homogeneous 
random potential in the approximation that the corresponding 
random Landau Hamiltonian on the Hilbert space ${\rm L}^2({\mathbbm{R}}^2) $ is restricted 
to the eigenspace of a single but arbitrary Landau level.
For a wide class
of $ {\mathbbm{R}}^2 $-homogeneous Gaussian random potentials we rigorously prove 
that the associated restricted integrated density of states 
is absolutely continuous with respect to the Lebesgue measure.
We construct explicit upper bounds on the resulting derivative, the restricted density
of states.
As a consequence, any given energy is seen to be almost surely not an eigenvalue of the restricted
random Landau Hamiltonian. \\[1ex] 
\noindent 
Physics and Astronomy Classification Scheme 2001: 71.70.Di, 71.23.-k, 02.30.Sa
}
\end{minipage} 
%
\end{abstract}
%
%
%
\tableofcontents
%
\section{Introduction} 
In recent decades
considerable attention has been paid to 
the physics of quasi-two-dimensional 
electronic structures.\cite{AnFoSt82,vKl86,KuMeTi88,Huc95,StTsGo99}
Some of the occurring phenomena, like the integer quantum Hall effect,\cite{Jan94} 
are believed to be microscopically explainable in terms of a Fermi gas
of non-interacting electrically charged particles in two dimensions subject to a
perpendicular constant magnetic field
and a static random potential.
For these phenomena it should therefore be sufficient to study
a single  non-relativistic spinless particle 
on the Euclidean plane $ {\mathbbm{R}}^2 $
modeled by the \emph{random Landau Hamiltonian}, 
which is informally given by 
\begin{equation}\label{eq:randLand}
  H\big(V^{(\omega)}\big) := H(0) + V^{(\omega)}.
\end{equation}
As a random Schr{\"o}dinger operator it
acts on the Hilbert space ${\rm L}^2({\mathbbm{R}}^2)$ of 
Lebesgue square-integrable 
complex-valued functions on the plane $ {\mathbbm{R}}^2 $.
For any realization $ \omega \in \Omega $ of the randomness
the potential $ V^{(\omega)} $ mimics the disorder present in a real sample. 
Throughout, we will assume that $ V $ is homogeneous on the average with respect to 
Euclidean translations of $ {\mathbbm{R}}^2 $.
%
The unperturbed part in (\ref{eq:randLand}) is the \emph{Landau Hamiltonian}.
It represents the kinetic energy of the particle and is informally  
given (in the symmetric gauge) by the differential expression
\begin{equation}\label{Eq:LandauHam}
H(0) :=  \frac{1}{2} \left[ %
                         \left(i \frac{\partial}{\partial x_1} -%
                             \frac{B}{2}x_2\right)^2 +%
                         \left(i \frac{\partial}{\partial x_2} +%
                             \frac{B}{2}x_1\right)^2 %
                       \right]
      = \frac{B}{2} \, \sum_{\ll=0}^{\infty} \left(2 \ll + 1\right) P_\ll,
\end{equation}
in physical units where the mass and the charge of the particle, and Planck's constant divided by
$ 2 \pi $ are all equal to one.
Moreover, $ B > 0 $ denotes the strength of the magnetic field and 
$ i = \sqrt{-1} $ stands for the imaginary unit.
The second equality in (\ref{Eq:LandauHam}) 
is the spectral resolution of $ H(0) $. It dates back to Fock\cite{Foc28} 
and Landau.\cite{Lan30}
The energy eigenvalue $(\ll + 1/2) B$ is called 
the $\ll^{\rm th}$ \emph{Landau level} and the corresponding orthogonal eigenprojection $P_\ll$ is an
integral operator with continuous complex-valued kernel (in other words: position representation)
\begin{equation}\label{eq:projektor}
P_\ll(x,y) := \frac{B}{2\pi}
  \exp\left[ i \frac{B}{2}  (x_2 y_1 - x_1 y_2) - \frac{B}{4} |x-y|^2\right] 
  {\rm L}^{(0)}_\ll\left(\frac{B}{2}|x-y|^2\right).
\end{equation}
Here and in the following, 
$ \left| x - y \right|^2 := (x_1-y_1)^2 + (x_2-y_2)^2  $
denotes the square of the Euclidean distance between the points $ x = (x_1, x_2) \in {\mathbbm{R}}^2 $ and
$  y = (y_1, y_2) \in {\mathbbm{R}}^2 $.
Moreover, $ {\rm L}^{(k)}_\ll(\xi):= \sum_{j=0}^\ll (- 1)^j \, \binom{\ll + k}{\ll -j} \, \xi^j/j! \,$, with
$ \xi \geq 0 $ and $ k \in {\mathbbm{N}}_0 - \ll := \left\{ - \ll,- \ll + 1 , -\ll +2 , \dots \right\} $,
is a generalized Laguerre polynomial,
see Sec.~8.97 in Ref.~\onlinecite{GrRy}.
The diagonal $P_\ll(x,x)= B/2\pi$ is naturally interpreted as the degeneracy
per area of the $ \ll^{\rm th} $ Landau level.

A quantity of basic interest in the study of the random Landau Hamiltonian (\ref{eq:randLand})
is its \emph{integrated density of states} $  \nu(] - \infty, E[) $ as a function of the energy $ E \in \mathbbm{R} $. 
The underlying positive Borel measure $ \nu $ on the real line $ {\mathbbm{R}} $  
is called 
the \emph{density-of-states measure} of $ H(V) $.  
If the random potential is not only 
${\mathbbm{R}}^2$-homogeneous but also \emph{isotropic}, that is, if all 
finite-dimensional distributions associated with the probability measure $ {\mathbbm{P}} $ on $ \Omega $, 
which governs the randomness, are invariant also under in-plane rotations (with respect to the origin),
the density-of-states measure $ \nu $ can be decomposed according to
\begin{equation}\label{eq:decomp}
  \nu = \frac{B}{2\pi} \sum_{\ll=0}^\infty \widehat\nu_\ll, 
  \qquad \qquad 
  \widehat\nu_\ll(I):= \frac{2\pi}{B} \,  {\mathbbm{E}}\left[
    \left(P_\ll \indfkt{I}(H(V))P_\ll\right)(x,x)
  \right],\qquad  I \in  \mathcal{B}({\mathbbm{R}}),
\end{equation}
see Refs.~\onlinecite{BrHeLe89,BrHuLe93}, and references therein.
Here $  {\mathbbm{E}}( \cdot ) := \int_{\Omega} {\mathbbm{P}}({\rm d} \omega) \, (\cdot ) $ 
denotes the expectation induced by $  {\mathbbm{P}} $ and
$ \indfkt{I}(H(V^{(\omega)})) $ is the spectral projection operator of $ H(V^{(\omega)}) $ 
associated with the energy regime $ I \in  \mathcal{B}({\mathbbm{R}}) $. 
The contribution  $ \widehat\nu_\ll$ related to the \emph{Landau-level index} $ \ll $ 
is a probability measure
on the Borel sets $ \mathcal{B}({\mathbbm{R}}) $ in the real line $ {\mathbbm{R}} $. 
It is actually independent of $x\in{\mathbbm{R}}^2$ due to the homogeneity of $ V $.

In the limit of a strong magnetic field,
the spacing $ B $ between successive Landau levels approaches 
infinity and the magnetic length $B^{-1/2}$ tends to zero. 
Therefore, the effect of so-called level mixing should be negligible if either the strength of the
random potential $ V $, typically given by the square root of its single-site variance 
$ {\mathbb{E}}\left[ V(0)^2 \right] - \left({\mathbb{E}}\left[ V(0) \right]\right)^2 $, 
is small compared to the level spacing or if the (smallest) correlation length
of $ V $ is much larger than the magnetic length. In both cases $\widehat\nu_\ll(I) $ should be well approximated
by $ 2\pi \, {\mathbbm{E}}\left[
\left(P_\ll \indfkt{I}(P_\ll H(V) P_\ll) P_\ll\right)(x,x)\right] / B $.
Indeed, this approximation is exact\cite{BrHeLe91} if $V$ is a spatially constant random 
potential $ x \mapsto V^{(\omega)}(0) $.
Since the first part of the $ \ll^{\rm th} $ \emph{restricted random Landau Hamiltonian},
$P_\ll H(V) P_\ll = (\ll + 1/2) B P_\ll + P_\ll V P_\ll$, causes only a shift in the energy,
one may equivalently study the probability
measure
\begin{equation}\label{defRIDOS}
   \nu_\ll(I):= \frac{2\pi}{B} \, {\mathbbm{E}} \left[\left(P_\ll \indfkt{I}(P_\ll V P_\ll)P_\ll\right)(x,x)\right],
   \qquad I \in \mathcal{B}({\mathbbm{R}}). 
\end{equation}
We call it the $ \ll^{\rm th} $ \emph{restricted density-of-states measure}
and its distribution function $ E \mapsto \nu_\ll (] - \infty , E [) $ the
$ \ll^{\rm th} $ \emph{restricted integrated density of states}.
Again, they are independent of $x\in{\mathbbm{R}}^2$ due to the homogeneity of $ V $.
From the physical point of view most interesting is the restriction 
to the \emph{lowest} Landau level,
corresponding to $\ll = 0$.
If the magnetic field is strong enough, all particles may be accommodated in the 
lowest level without conflicting with Pauli's exclusion principle,
since the degeneracy (per area) $B/2\pi$ increases with $ B $.
Up to the energy shift $ B/2 $, the measure $ B \nu_0 / 2 \pi$ 
should then be a good approximation to $\nu$, 
since the effects of higher Landau levels are negligible 
if $B$ is large compared to the strength of the
random potential, see Prop.~1 in Ref.~\onlinecite{MaPu92} in case of a 
Gaussian random potential.

Neglecting effects of level mixing by only dealing with the sequence of 
restricted operators $\left( P_\ll V P_\ll \right)_{l\in{\mathbbm{N}}_{0}}$
is a simplifying approximation which is often made.
The interest in these operators 
relates to the existence of pure-point components 
in their spectra\cite{DoMa95,DoMa96,DoMa97,Scr99} and, what is simpler, to properties of their restricted  
density-of-states 
measures $ ( \nu_\ll ) $.
The aim of the present paper is to supply conditions under which $ \nu_\ll $ is absolutely continuous with respect to the
Lebesgue measure.
Actually, in the physics literature
this differentiability of the restricted integrated density of states
$ \nu_\ll\left(]-\infty , E[\right)  $ with respect to $ E $ is usually taken for granted  
so that one deals from the outset with its derivative, 
the $ \ll^{\rm th} $~\emph{restricted density of states}
\begin{equation}\label{eq:dos}
  E \mapsto w_\ll(E) 
  := \frac{{\rm d}\nu_\ll\left(]- \infty , E[\right)}{{\rm d}E} 
  = \frac{\nu_\ll( {\rm d} E)}{{\rm d}E},
\end{equation} 
see, for example, 
Refs.~\onlinecite{AnFoSt82,vKl86,KuMeTi88,Huc95,StTsGo99,Jan94,BrHeLe89,BrHeLe91,Weg83,BeGrIt84,KlPe85,BenCha86,Sal87,Ape87,BrHeLe90,BoBrLe97}.
Due to the involved averaging, the disorder is indeed often believed to
broaden each Landau level to a \emph{Landau band} in such a way that 
the resulting restricted integrated density of states is sufficiently smooth.
Example~1 below however, which is taken from Ref.~\onlinecite{BrHeLe91}, illustrates 
that this belief is wrong without further assumptions.
It even shows that for any given $\ll \geq 1$ it may happen that there is no broadening 
at all so that the operator $P_\ll V  P_\ll$ is zero 
almost surely (although $V$ is non-zero) and hence $ \nu_\ll$ is singular.
For the formulation of the example we need some preparations.
Without losing generality, we will always 
assume that the homogeneous (but not necessarily isotropic)
random potential $ V $ 
has  zero mean, ${\mathbbm{E}}[\, V(0)] = 0 $.
The \emph{variance} $ \sigma_\ll^2 $ of $ \nu_\ll $ is then given by\cite{BrHeLe91}
  \begin{equation}\label{Eqbreite}
     \sigma_\ll^2 :=  \int_{{\mathbbm{R}}} \nu_\ll({\rm d} E)\, E^2 =
     \frac{2\pi}{B} \, {\mathbbm{E}} \left[ \left(P_\ll V P_\ll \right)^2(0,0) \right] 
     = \frac{2\pi}{B} \, \left(P_\ll C P_\ll\right)(0,0)  \leq C(0),
  \end{equation}
where $x \mapsto C(x) := {\mathbbm{E}}\left[V(x) V(0) \right]$ is 
the \emph{covariance function} of $ V $. 
When sandwiched between 
two projections, 
$ C $ is understood as a (bounded) multiplication operator acting on $ {\rm L}^2({\mathbbm{R}}^2) $.
The standard deviation $ \sigma_\ll := \sqrt{ \sigma_\ll^2} $ may physically be interpreted as the 
width of the $ \ll^{\rm th} $ Landau band.
We note that the width  $ \sigma_0 $ of the lowest Landau band
is always strictly positive, provided that the covariance function is continuous and obeys 
$ C(0) > 0 $. This follows from the formula 
$ \sigma_\ll^2 = \int_{{\mathbbm{R}}^2} \widetilde C({\rm d}^2k) \, \exp\big(- |k|^2 /2 B  \big)  
\big[ {\rm L}^{(0)}_\ll\left(|k|^2/2 B \right)\big]^2 $.
Here the so-called \emph{spectral measure} $ \widetilde C $  which, 
according to the Bochner-Khintchine theorem (Thm.~IX.9 in
Ref.~\onlinecite{ReSi80}), is 
the unique finite positive (and even) Borel measure on $ {\mathbbm{R}}^2 $ yielding the Fourier representation 
$ C(x) = \int_{{\mathbbm{R}}^2}  \widetilde C({\rm d}^2k) \, \exp\left(i k \cdot x\right) $ 
where $ k \cdot x := k_1 x_1 + k_2 x_2 $ denotes the standard scalar product 
on $ {\mathbbm{R}}^2 $.\\

\noindent
{\bf Example~1.}~
  If $V$ possesses the oscillating covariance 
  function~~$C(x)=C(0)\, J_0\left(\sqrt{2} |x| /\tau \right)$, where $\tau > 0$
  and $J_0$ is the Bessel function of order zero,\cite{GrRy} then
  \begin{equation}
     \sigma_\ll^2 = C(0) \, \exp\left(- \frac{1}{B \tau^2}\right)
    \left[{\rm L}^{(0)}_\ll\left(\frac{1}{B \tau^2}\right)\right]^2.
  \end{equation}
  Choosing the squared length ratio $1/(B \tau^2)$ equal to a zero of ${\rm L}^{(0)}_\ll$, 
  which exists if $\ll \geq 1$, 
  one achieves
  that $ \sigma_\ll^2=0$. Chebyshev's inequality then implies that $ \nu_\ll$ is
  Dirac's point measure at the origin, 
  informally $ w_\ll(E) = \delta(E) $. 
  Therefore, $P_\ll V^{(\omega)} P_\ll = 0$ for $\mathbbm{P}$-almost 
  all $\omega \in \Omega$.\\

\noindent
In the present paper we  
provide conditions under which exotic situations
as in Example~1 cannot occur. 
More precisely, we prove that (\ref{eq:dos}) indeed defines $ w_\ll $ as 
a bounded probability density for a wide class of \emph{Gaussian} random potentials, see Theorem~1 and Theorem~2 
below. Moreover, we construct explicit upper bounds on $ w_\ll $
for these potentials.
As an implication,
we prove that for any $ B > 0 $ and the class of Gaussian random potentials considered, 
any given energy $ E \in {\mathbbm{R}} $ is 
almost surely not an eigenvalue of the operator $ P_\ll V P_\ll $. 
In particular, these Gaussian random potentials completely
lift the infinite degeneracy of the Landau-level energy (here shifted to zero)
for any strength of the magnetic field. This stands in contrast to situations 
with random point impurities considered in Refs.~\onlinecite{BeGrIt84,DoMa97,PuSc97,Erd98,Scr99}.\\

The present paper was partially motivated by results of Ref.~\onlinecite{HLMW00}
where
the (unrestricted) density-of-states measure $ \nu $ of the random Landau Hamiltonian is proven to be absolutely continuous
with a locally bounded density for a certain class of random potentials (including the Gaussian ones considered in Theorem~1) 
and where any given 
energy $ E \in {\mathbbm{R}} $ is shown to be almost surely not an eigenvalue of $ H(V) $.
While absolute continuity of $ \nu $ immediately implies  by (\ref{eq:decomp}) that of $ \widehat \nu_\ll $ 
for all $ \ll \in {\mathbbm{N}}_0 $ (if $ V $ is isotropic), in itself
it does not imply that of $ \nu_\ll $.
\section{The density of states of a single broadened Landau level}\label{Ch:5}
\subsection{The restricted random Landau Hamiltonian and its integrated density of states}
Let $ \left\| F \right\| := \sup \big\{ \left| \langle \varphi , F \varphi \rangle \right|  
    \, : \, \varphi \in {\rm L}^2({\mathbbm{R}}^2) \, , \,  \langle \varphi , \varphi \rangle = 1 \big\} 
  < \infty $ denote the (uniform) norm of a self-adjoint bounded operator $F$ acting on the 
Hilbert space ${\rm L}^2({\mathbbm{R}}^2)$.
The restriction $P_\ll F P_\ll$ of $F$ 
to the eigenspace  $ P_\ll{\rm L}^2({\mathbbm{R}}^2) \subset {\rm L}^2({\mathbbm{R}}^2) $ 
corresponding to the $ \ll^{\rm th} $ Landau level is an integral operator with kernel
$(x,y) \mapsto \left(P_\ll F P_\ll\right)(x,y) := B\,\langle \psi_{\ll,x} \, , F \, \psi_{\ll,y} \rangle /2\pi$
which is jointly continuous thanks to the continuity of the usual scalar product $ \langle \cdot , \cdot \rangle $
on $ {\rm L}^2({\mathbbm{R}}^2) $ and the strong 
continuity of the mapping 
${\mathbbm{R}}^2 \ni x \mapsto \psi_{\ll,x} \in P_\ll{\rm L}^2({\mathbbm{R}}^2)$.
Here, the two-parameter family of normalized, complex-valued functions (``coherent states'') is 
defined by
\begin{equation}\label{defpsi}
  y \mapsto \psi_{\ll,x} (y):= \sqrt{\frac{2\pi}{B}}\, P_\ll(y,x), \qquad 
  x\in{{\mathbbm{R}}}^2,\qquad  \langle\psi_{\ll,x} \, , \, \psi_{\ll,x}\rangle = 1.
\end{equation}

Let $ (\Omega , {\mathcal{A}}, {\mathbbm{P}}) $ be a complete probability
space.
By a \emph{random potential} we mean a random field
$ V: \Omega \times {{\mathbbm{R}}}^2 \to {{\mathbbm{R}}}$, 
$(\omega, x) \mapsto V^{(\omega)}(x) $ which is jointly measurable with 
respect to the sigma-algebra $ {\mathcal{A}} $ of event sets in $ \Omega $ and the 
sigma-algebra $ {\mathcal{B}}({\mathbb{R}}^2) $ of Borel sets in the 
Euclidean plane ${\mathbb{R}}^2 $.

The next proposition provides conditions under which 
the (in general unbounded) integral operator 
$P_\ll V P_\ll$, the (shifted) $ \ll^{\rm th} $ \emph{restricted random Landau Hamiltonian},
 is almost surely essentially self-adjoint on the Schwartz space
${\mathcal{S}}({{\mathbbm{R}}}^2)$ of arbitrarily often differentiable 
complex-valued functions on ${{\mathbbm{R}}}^2 $ with rapid decrease (Def. on p.~133 in Ref.~\onlinecite{ReSi80}).\\ 

%
\noindent
{\bf Proposition 1.}~ {\itshape
  Let $V$ be an ${{\mathbbm{R}}}^2$-homogeneous random potential and assume
  there
  exists a constant $M< \infty$ such that ~${\mathbbm{E}}\left[ |V(0)|^{2k}\right] \leq (2 k)!\, M^{2k} $
  for all $k \in \mathbbm{N}$. Then
  for all $\ll \in {\mathbb{N}}_0$ it holds:
  \begin{enumerate}
  \item[(1)]
    The \emph{restricted operator}
    $P_\ll V^{(\omega)} P_\ll$ is
    essentially self-adjoint on ${\mathcal{S}}({{\mathbbm{R}}}^2)$
    for all $ \omega $ in some subset $ \Omega_0 \in \mathcal{A} $ of\ $ \Omega $ with
    full probability, ${\mathbbm{P}}(\Omega_0) = 1 $. 
  \item[(2)] 
    The mapping $\Omega_0 \ni \omega \mapsto P_\ll V^{(\omega)} P_\ll$ is measurable
    in the sense of Def.~V.1.3 in Ref.~\onlinecite{CaLa90}. 
  \item[(3)]
    The \emph{restricted density-of-states measure} $ \nu_\ll$, defined in~(\ref{defRIDOS}),
    is a probability measure on the sigma-algebra $ {\mathcal{B}}({\mathbbm{R}}) $ of Borel sets in the real line. 
    Moreover, the
    following (weak) operator identity holds
    \begin{equation}\label{Schur}
      {\mathbbm{E}}\left[ P_\ll \indfkt{I}(P_{\ll} V P_{\ll}) P_\ll \right] =  \nu_{\ll}(I) \,  P_\ll,
      \qquad I \in \mathcal{B}({{\mathbbm{R}}}).
    \end{equation}
  \end{enumerate} 
}

\noindent
{\bf Remark.}~ 
As far as we know, the operator identity (\ref{Schur}) 
for general $ {\mathbb{R}}^2 $-homogeneous $ V $ 
and general $ \ll \in {\mathbb{N}}_0 $
was first shown in Ref.~\onlinecite{KlPe85},
also see Ref.~\onlinecite{BrHeLe90}.\\

\noindent
{\bf Proof of Proposition~1.}~ 
  ~(1)~ The assumed limitation on the growth of the even moments ${\mathbbm{E}}\left[ |V(0)|^{2k}\right] $ 
  as a criterion for the almost-sure essential self-adjointness of $ P_\ll V P_\ll $ 
  is taken from the proof  of Thm.~2.1 in Ref.~\onlinecite{DoMa95}, which is based
  on Nelson's analytic-vector theorem (Thm.~X.39 in Ref.~\onlinecite{ReSi75}), also 
  see Ref.~\onlinecite{BrHuKiLe95}.\\[1ex]
\indent
 (2)~ By truncating the large fluctuations of the random potential we construct a sequence
    of restricted random operators 
    $\big( P_\ll V_n^{(\omega)} P_\ll \big)_{n \in \mathbbm{N}}$, 
    where 
    $
    V_n^{(\omega)}(x) := V^{(\omega)}(x) \, \Theta\left(n - \left|V^{(\omega)}(x)\right|\right) 
    $ is bounded and measurable for all $ n $. Here $ \Theta := \indfkt{] 0 , \infty [ } $ 
    denotes Heaviside's unit-step function.
    For $ {\mathbbm{P}} $-almost all $ \omega \in \Omega $ we have the strong convergence 
    $ P_\ll V_n^{(\omega)} P_\ll \varphi \to P_\ll V^{(\omega)} P_\ll \varphi $ as $n \to \infty $ 
    for all $ \varphi \in {\mathcal{S}}({{\mathbbm{R}}}^2)$. Consequently, Thm.~VIII.25 in Ref.~\onlinecite{ReSi80}
    implies that  
    the sequence converges towards $P_\ll V^{(\omega)} P_\ll$ in 
    the strong resolvent sense implying that the
    latter operator is also measurable thanks to Prop.~V.1.4 in Ref.~\onlinecite{CaLa90}.\\[1ex]
\indent
 (3)~ Since $ 0 \leq \left\langle \psi_{\ll,x} \, , \, \indfkt{I}\!\left( P_\ll V^{(\omega)} P_\ll \right) \, \psi_{\ll,x} \right\rangle
             \leq  \left\langle \psi_{\ll,x} \, , \, \indfkt{{\mathbbm{R}}}\!\left(P_\ll V^{(\omega)} P_\ll \right) \, \psi_{\ll,x} \right\rangle
  = 1 $ for all $ x \in {\mathbbm{R}}^2 $, all $ I \in {\mathcal{B}}({\mathbbm{R}}) $, and
  $ \mathbbm{P} $-almost all $ \omega \in \Omega $, the r.h.s. of~(\ref{defRIDOS}) indeed defines
  a probability measure on $ {\mathcal{B}}({\mathbbm{R}}) $. 
  For the proof of~(\ref{Schur}) we introduce the family of   
    \emph{magnetic translation operators}
    $\left\{T_x\right\}_{x \in {\mathbbm{R}}^2} $ which are unitary on 
    $ {\rm L}^2({{\mathbbm{R}}}^2) $ and defined by
    \begin{equation}\label{eq:magntrans}
      \left( T_x \varphi \right)(y) := \exp\left[ i \frac{B}{2} \left(x_1 y_2 - x_2 y_1
          \right) \right] \, \varphi(y-x), \qquad \varphi \in  {\rm L}^2({{\mathbbm{R}}}^2).
    \end{equation}
    They 
    constitute an irreducible representation
    of the Heisenberg-Weyl group on 
    $ P_\ll {\rm L}^2({{\mathbbm{R}}}^2)$. \cite{KlPe85} 
    Since $V$ is ${\mathbbm{R}}^2$-homogeneous, it follows that
    \begin{equation}
      T_x^\dagger \,  
      {\mathbbm{E}}\left[ P_\ll \indfkt{I}(P_{\ll} V P_{\ll}) P_\ll \right] 
      T_x =
      {\mathbbm{E}}\left[ P_\ll \indfkt{I}(P_{\ll} V P_{\ll}) P_\ll \right]
    \end{equation}
    for all $x \in {{\mathbbm{R}}}^2$ and all $I \in \mathcal{B}({{\mathbbm{R}}})$.
    A suitable variant of Schur's lemma (Prop.~4 of {\S}3, Ch.~5 in Ref.~\onlinecite{BaRa77}) 
    then gives the claimed result. \qed \\

\noindent In Theorem~1 and Theorem~2 below, we will assume that $ V $ is a 
Gaussian random potential in the sense of\\

\noindent
{\bf Definition 1.}~ A \emph{Gaussian random potential} is a  
Gaussian\cite{Adl81,Lif95}  random field $ V $ which is 
${\mathbb{R}}^2 $-homogeneous, has zero mean,
$ {\mathbbm{E}}\left[ V(0) \right] = 0 $, and is characterized by a covariance function
$ {\mathbbm{R}}^2 \ni x \mapsto C(x) := {\mathbbm{E}}\left[ V(x) V(0) \right] $ which is continuous
at the origin where it obeys $ 0 < C(0) < \infty $.\\

\noindent
{\bf Remarks.}~ ~(1)~ Our continuity requirement for the covariance function $ C $
of a Gaussian random potential implies 
that $ C $ is uniformly continuous and bounded by $ C(0) $. 
Consequently, there exists a separable version 
of $ V $ which is jointly measurable, see Thm.~3.2.2 in Ref.~\onlinecite{Fer75}.
When speaking about a Gaussian random potential, we will tacitly assume
that only this version is dealt with.\\[1ex]
\indent 
(2)~ A Gaussian random potential fulfills the assumption of Proposition~1 
with $M = \sqrt{C(0)}$ by the usual ``Gaussian combinatorics'' 
to be found, for example, 
in Lemma~5.3.1 of Ref.~\onlinecite{Adl81}.


\subsection{Existence and boundedness of the restricted density of states}
Wegner estimates \cite{Weg81} have turned out to be an efficient tool for 
proving the absolute continuity of density-of-states measures for certain random operators and 
for deriving upper bounds on their respective Lebesgue densities. 
One method to derive estimates of this genre uses \emph{one-parameter spectral averaging}.
It provides upper bounds on the averaged spectral projections 
of a self-adjoint operator which is
perturbed by a bounded positive operator with fluctuating coupling strength.
The abstract version of such an averaging, which we will use, is due to 
Combes and Hislop.\cite{CoHi94}
It is rephrased as\\

\noindent
{\bf Lemma~1.}~ 
{\itshape
   Let $K$, $L$, and $M$ be three self-adjoint operators acting
  on a Hilbert space $\mathcal{H}$ with scalar product $ \langle \cdot , \cdot \rangle $.
  Moreover, let $K$ and $M$ be bounded such that
  ~$ \kappa := \inf_{\varphi \in {\mathcal{H}}, \, K \varphi {\neq} 0 } \,  
    \langle \varphi \, , \, M  \, \varphi \rangle / 
    \langle \varphi \, , \, K^2 \,  \varphi \rangle > 0 $ is strictly positive.
  Finally, let $ g \in {\rm L}^\infty(\mathbbm{R})$ 
  be a Lebesgue-essentially bounded function on the real line,
  $ \left\|g\right\|_\infty := {\rm ess\,sup}_{\xi \in {\mathbbm{R}}}\, \left| g(\xi) \right| < \infty $. 
  Then the inequality
  \begin{equation}\label{LSM}
    \int_{{\mathbbm{R}}} \! {\rm d}\xi  \, \, \left| g(\xi) \right| \, \langle \varphi \, , \, 
    K \, \indfkt{I}(L + \xi M) \, K \, \varphi \rangle 
    \leq |I| \, \frac{\left\|g\right\|_\infty}{\kappa} \, \langle \varphi , \varphi \rangle
  \end{equation}
  holds for all $\varphi \in\mathcal{H}$ and all $I \in \mathcal{B}(\mathbbm{R})$. }\\

\noindent
{\bf Proof.}~ See Cor.~4.2 in Ref.~\onlinecite{CoHi94} and  
     Lemma~3.1 in Ref.~\onlinecite{HLMW00}. \qed \\

\noindent
If one focuses only on the absolute continuity of the measure $ \nu_\ll$  without
aspiring after sharp upper bounds on the resulting Lebesgue density, 
a straightforward application of Lemma~1 \nopagebreak[9]
yields the following\\

\noindent
{\bf Theorem 1.}~  
{\itshape
  Let $V$ be a Gaussian random potential in the sense of Definition~1. Suppose that
  there exists a finite 
  signed Borel measure $\mu$ on ${{\mathbbm{R}}}^2$ such that the covariance function $ C $ 
  of $ V $ obeys
  \begin{equation}\label{vorranz}
    0 \leq C_\mu(x) := \int_{{{\mathbbm{R}}}^2} \! \mu({\rm d}^2y) \, C(x-y)  < \infty , \qquad\qquad
    \int_{{{\mathbbm{R}}}^2} \! \mu({\rm d}^2y) \, C_\mu(y) = 1
  \end{equation}
  for all $ x \in {{\mathbbm{R}}}^2 $.
  Then the $\ll^{\rm th}$ restricted density-of-states measure $ \nu_\ll$ is absolutely 
  continuous with respect to the Lebesgue measure 
  and the resulting Lebesgue probability density $  w_\ll $, the \emph{restricted density of states}, 
  is uniformly bounded according to
  \begin{equation}\label{ranzgauss}
    w_\ll(E) := \frac{\nu_\ll({\rm d} E)}{{\rm d} E} \leq \frac{1}{\sqrt{2\pi} \,
      \left\|P_\ll  C_\mu  P_\ll \right\|}
  \end{equation} 
  for Lebesgue-almost all energies $ E \in {{\mathbbm{R}}} $.
  Moreover, any given  $ E  \in {{\mathbbm{R}}} $ is not an eigenvalue of
  ~$ P_\ll \, V^{(\omega)} \, P_\ll $~ for $ \mathbbm{P} $-almost all $ \omega \in \Omega $. 
}\\

\noindent
{\bf Remarks.}~
 (1)~ The second equality in (\ref{vorranz}) is just a convenient normalization of $ \mu $.
  The measure $ \mu $ allows one to optimize the 
  upper bound in (\ref{ranzgauss})
  (see Example~2 below) 
  as well as
  to apply Theorem~1 to
  Gaussian random potentials with certain covariance functions $ C $ taking on
  also negative values.
  One such example is 
  $ C(x) = C(0)\, \exp\left[ - |x|^2/(2\tau^2) \right] \, \big[ 1 - 7 |x|^2/(16 \tau^2) + |x|^4/(32 \tau^4) \big]$
  with arbitrary length scale $ \tau > 0$.
  This may be seen by choosing the Gaussian measure
  $ \mu\left({\rm d}^2 x\right) = {\rm d}^2 x \, N \, \exp\left[ - | x |^2/(8 \tau^2) \right] $ 
  with a suitable normalization factor $ N > 0 $.
  Of course, for the oscillating covariance function of
  Example~1 no $ \mu $ exists yielding the positivity condition (\ref{vorranz}).\\[1ex] 
\indent 
(2)~ We note that 
   $ 0 < \langle \psi_{\ll,0} \, , C_\mu \, \psi_{\ll,0} \rangle \leq \left\| P_\ll C_\mu P_\ll \right\| < \infty $.
   The first (strict) inequality follows from the assumptions on~$ C $ in Definition~1, Eq.~(\ref{vorranz}), 
   and the explicit
   form (\ref{defpsi}) of $ \psi_{\ll,0} $.\\

\noindent
{\bf Proof of Theorem~1.}~
  The proof consists of two parts. 
  In the first part, we use the fact that the Gaussian random potential $ V $ admits 
  a one-parameter decomposition into a standard Gaussian random variable 
  $ \lambda $ and a non-homogeneous zero-mean
  Gaussian random field $ U $ which are defined by
  \begin{equation}\label{eq:decomposition}
    \lambda^{(\omega )} := \int_{{{\mathbbm{R}}}^2}\!\mu ({\rm d}^d y) \, V^{(\omega )}(y), 
    \qquad\qquad
     U^{(\omega)}(x) := V^{(\omega)}(x) -  \lambda^{(\omega )} \, C_\mu(x).
  \end{equation}
  The positive bounded function $ C_\mu $ is defined in (\ref{vorranz}). 
  Since $ {\mathbbm{E}}\left[  \lambda \, U(x) \right] = 0 $ for all $ x \in {\mathbbm{R}}^2 $,
  $ \lambda $ and $ U $ are stochastically independent.
  We now multiply both sides of (\ref{Schur}) from the left and right 
  by $K := ( P_\ll C_\mu P_\ll )^{1/2} $ and take the quantum-mechanical 
  expectation with respect to an arbitrary non-zero
  $\varphi \in P_\ll {\rm L}^2({{\mathbbm{R}}}^2)$ to 
  obtain
  \begin{align}
    & \nu_\ll(I) \, \, \langle \varphi \, , K^2 \varphi \rangle
      = {\mathbbm{E}}\Big[ \langle\varphi \, , \, K\, 
       \indfkt{I}(P_{\ll} V P_{\ll})\, K \varphi\rangle \Big] \notag \\
    & \qquad   =  {\mathbbm{E}} \left[ \int_{{\mathbbm{R}}}\! {\rm d} \xi \, 
       \frac{ {e}^{-\xi ^2/2}}{\sqrt{2\pi}} \, 
       \langle \varphi \, , \, K \, 
       \indfkt{I}(P_{\ll} U P_{\ll} + \xi K^2) \, K \, \varphi \rangle \right]
     \leq \left|I\right|  \, \frac{\langle \varphi \, , \, \varphi \rangle}{\sqrt{2\pi}}.
     \label{eq:vorrechnen}
  \end{align}
  For the second equality we used the one-parameter decomposition (\ref{eq:decomposition}) 
  and the stochastic
  independence of $ \lambda $ and $ U $. 
  The Lebesgue integral in (\ref{eq:vorrechnen}), which constitutes a partial averaging, 
  is then bounded with the help of (\ref{LSM})
  uniformly in $ \omega $. 
  The absolute continuity of $ \nu_\ll $ with
  respect to the Lebesgue measure is now a consequence of (\ref{eq:vorrechnen}) and
  the Rad\'on-Nikod\'ym theorem.
  Minimizing the upper bound on $ \nu_\ll(I) $, coming from (\ref{eq:vorrechnen}), with respect to 
  $\varphi \in P_\ll {\rm L}^2({{\mathbbm{R}}}^2)$ 
  yields the claimed inequality~(\ref{ranzgauss}).
  
  In the second part, we note that (\ref{Schur}) implies the equivalence: ~$  \nu_\ll $ 
  has no pure points, that is, $ \nu_\ll (\{E\}) = 0 $
  for all $ E \in {\mathbbm{R}} $,
  if and only if 
  $ {\mathbbm{E}}\left[ \langle \varphi , \, 
    \indfkt{\{E\}}(P_\ll V P_\ll) \, \varphi \rangle
  \right] = 0 $ for all $ E \in {\mathbbm{R}} $ and all
  $ \varphi \in  P_\ll  {\rm L}^2({{\mathbbm{R}}}^2) $.
  Given an orthonormal basis $ \left\{ \varphi_k \right\}_{k \in {\mathbbm{N}}} $ in
  $ P_\ll  {\rm L}^2({{\mathbbm{R}}}^2) $, there hence exists for every $ k \in {\mathbbm{N}} $
  some $ \Omega_k \in {\mathcal{A}} $
  with $ {\mathbbm{P}}(\Omega_k) = 1 $ such that $  \langle \varphi_k , \, 
  \indfkt{\{E\}}(P_\ll V^{(\omega)} P_\ll)  \varphi_k \rangle = 0 $ for
  all $ \omega \in \Omega_k $. As a consequence, 
  $ \indfkt{\{E\}}(P_\ll V^{(\omega)} P_\ll)  = 0 $ for all 
  $  \omega \in \cap_{ k \in {\mathbbm{N}}} \Omega_k $, 
  hence $ {\mathbbm{P}} $-almost all $ \omega \in \Omega $.
 \qed \\

\noindent
{\bf Remarks.}~
(1)~ If the spectral measure $ \widetilde C $ has 
a (positive) Lebesgue density, 
$ C $ admits the representation 
$ C(x) = \int_{{\mathbbm{R}}^2} \! {\rm d}^2y \,\, \gamma(x+y) \, \gamma(y) $ 
with some $ \gamma \in {\rm L}^2({\mathbbm{R}}^2) $. 
If furthermore there exists some $ f \in {\rm L}^2({\mathbbm{R}}^2) $ 
with $ \langle f , f \rangle = 1 $ such that
$ 0 \leq u(x) := \int_{{{\mathbbm{R}}}^2} \! {\rm d}^2y \, \, f(x+y) \, \gamma(y) < \infty $
and $ u \neq 0 $, one may replace $ C_\mu $ in (\ref{ranzgauss}) by $ u $ to obtain 
another upper bound
on $ w_\ll $ for the given Gaussian random potential $ V $.
Roughly speaking, the idea is to write $ V $  
as $ V^{(\omega)}(x) = \int_{{\mathbbm{R}}^2} {\rm d}^2y \, \gamma(x+y) \, W^{(\omega)}(y) $,
where $ W $  is the standard delta-correlated (generalized) Gaussian random field on $ {\mathbbm{R}}^2 $
informally characterized by $ {\mathbbm{E}} \left[\, W(x) \right] = 0 $ and 
$ {\mathbbm{E}} \left[ \, W(x) W(y) \right] = \delta(x-y) $. 
The Gaussian random potential $ V $  hence admits 
a one-parameter decomposition into the standard
Gaussian random variable $ \lambda^{(\omega)} := \int_{{\mathbbm{R}}^2} {\rm d}^2y \, f(y) \,  W^{(\omega)}(y) $ 
and the non-homogeneous
Gaussian random field $ U^{(\omega)}(x) := V^{(\omega)}(x) -  \lambda^{(\omega )} \, u(x) $,
which is stochastically independent of $ \lambda $.\\[1ex]
\indent  
(2)~ The essential ingredients of the above proof are the 
operator identity (\ref{Schur}) and the fact 
that the (not necessarily Gaussian)
random potential admits a one-parameter
decomposition\cite{FiHu97b,HLMW00} $ V^{(\omega)}(x) = U^{(\omega)}(x) + \lambda^{(\omega)} u(x) $ into a positive
function $ u $, 
a random field $ U $, and a random variable $ \lambda $ whose conditional
probability measure with respect to the sub-sigma-algebra generated 
by the family of random variables $ \left\{ U(x) \right\}_{x \in {\mathbbm{R}}^2} $ has a bounded Lebesgue density $ \varrho $.
Following the lines of reasoning of the above proof, the restricted density of states may then be shown to
be bounded according to
\begin{equation}\label{eq:allg}
  w_\ll(E) = \frac{\nu_\ll({\rm d}E)}{{\rm d}E} \leq \frac{\| \varrho \|_\infty}{ \| P_\ll u P_\ll \|}
\end{equation}
for Lebesgue-almost all $ E \in {\mathbbm{R}} $. Moreover, any given 
$ E \in {\mathbbm{R}} $ is not an eigenvalue of 
$ P_\ll V^{(\omega)} P_\ll $ for $ {\mathbbm{P}} $-almost all $ \omega \in \Omega$.\\[1ex]
\indent 
(3)~ The energy-independent estimate (\ref{ranzgauss}) 
  is rather rough, because one expects $ w_\ll(E) $
  to fall off to zero for energies $ E $ approaching 
  the edges $ \pm \infty $ (if $ \sigma_\ll^2 > 0 $) of the almost-sure spectrum of $ P_\ll V P_\ll $.
  More precisely, in the present case of a Gaussian random potential $ V $ it follows from arguments in
  Ref.~\onlinecite{BrHeLe91} that the leading asymptotic behavior
  of the restricted integrated density of states 
  for $ \left| E \right| \to \infty $ is Gaussian according to 
  \begin{equation}\label{eq:tail}
    \lim_{E \to - \infty} \, \frac{\ln \nu_\ll (] - \infty , E [)}{E^2} 
    = 
    \lim_{E \to \infty} \, \frac{\ln \nu_\ll (] E , \infty [) }{E^2}
    =  - \frac{1}{2 \Gamma_\ll^2},
  \end{equation}
  where the \emph{decay energy} $ \Gamma_\ll $ is the solution of the maximization problem
  \begin{equation}\label{Def:Gamma}
    \Gamma_\ll^2 := \!\! \sup_{\substack{\varphi \in P_\ll{\rm L}^2({\mathbbm{R}}^2)
    \\ \langle \varphi , \varphi \rangle= 1}}  \gamma^2(\varphi), \quad 
  \gamma^2(\varphi) := {\mathbbm{E}}\left[ \langle \varphi, V \varphi \rangle^2 \right] = 
  \int_{{\mathbbm{R}}^2} \! \! \! \! {\rm d}^2 x  
  \int_{{\mathbbm{R}}^2} \! \! \! \! {\rm d}^2 y   \left| \varphi(x) \right|^2  \left| \varphi(y) \right|^2 
  C(x-y).
\end{equation} 
We recall from Ref.~\onlinecite{BrHeLe91} the inequalities $ \sigma_\ll^4/C(0) \leq \Gamma_\ll^2 \leq \sigma_\ll^2 $.\\

\noindent

\noindent
Gaussian random potentials with positive covariance functions nicely illustrate Theorem~1.\\

\noindent 
{\bf Example~2.}~
If $0 \leq C(x) < \infty$ for all $x \in {{\mathbbm{R}}}^2$, the optimal $ \mu $
in (\ref{ranzgauss}) belongs to the class of positive measures of the form 
$ \mu({\rm d}^2x) \, \gamma(\varphi) = {\rm d}^2x \, \left| \varphi(x) \right|^2 $ with 
$ \varphi \in P_\ll {\rm L}^2({\mathbbm{R}}^2) $, $\langle \varphi , \varphi \rangle = 1 $, and
$ \gamma(\varphi) $ as defined in (\ref{Def:Gamma}). Optimizing with respect to $ \varphi $ yields
\begin{equation}\label{eq:Cpos}
  w_\ll(E) \leq \frac{1}{\sqrt{2 \pi \Gamma_\ll^2}},
\end{equation}
where $ \Gamma_\ll $ is the decay energy of the restricted integrated density of states.
In particular, for the Gaussian covariance function 
~$ C(x) = \alpha^2 \,  \exp\big[- \left| x \right|^2/( 2 \tau^2) \big]/ (2 \pi \tau^2) $~
with correlation length $ \tau > 0 $ and single-site variance $ C(0) = \alpha^2/ (2 \pi \tau^2) > 0 $, 
one has explicitly\cite{BrHeLe91} 
\begin{equation}\label{eq:decayGauss}
  \Gamma_\ll^2 = \gamma^2(\varphi_{\ll,-\ll})  
  = \frac{\alpha^2}{2 \pi \tau^2} \, \left( \frac{B \tau^2}{B \tau^2 +2} \right)^{\ll + 1} \,
          {\rm P}_\ll\left(\frac{(B \tau^2 + 1)^2 +1}{(B \tau^2 + 1)^2 - 1}\right),
\end{equation}
where the maximizer $ \varphi_{\ll,-\ll} $ is given in (\ref{EqAngular}) below and 
$ {\rm P}_\ll(\xi) := ( 1 /\ll! \, 2^\ll ) 
({\rm d}^\ll / {\rm d}\xi^\ll  ) (\xi^2 - 1)^\ll $
is the $ \ll^{\rm th} $ Legendre polynomial.\cite{GrRy}\\  

\noindent
{\bf Remarks.}~ 
  (1)~ That the class of measures referred to in Example~2 contains indeed the optimal one,  
  derives from the Fourier representation
  \begin{equation}\label{eq:fourier}
    \langle \varphi \, , C_\mu \, \varphi \rangle = \int_{{\mathbbm{R}}^2} \!\!  \widetilde C\left({\rm d}^2k\right) \;
    \left( \int_{{\mathbbm{R}}^2} \!\! {\rm d}^2x \, \left| \varphi(x) \right|^2 {e}^{i k \cdot x} \right) \,
    \left( \int_{{\mathbbm{R}}^2} \!\! \mu\left( {\rm d}^2y\right) \, {e}^{-i k \cdot y} \right)
  \end{equation}
  valid for all $ \varphi \in {\rm L}^2({\mathbbm{R}}^2) $. 
  Since $ \widetilde C $ is positive, 
  the claim follows from (\ref{eq:fourier}) with the help of the Cauchy-Schwarz inequality
  and the positivity of $ C $.\\[1ex]
  \indent 
  (2)~ In the physics literature one often considers the limit of a  \emph{delta-correlated}
 Gaussian random potential informally characterized by  $C(x) = \alpha^2 \, \delta(x) $ 
 with some $ \alpha > 0 $. 
 It emerges from the Gaussian random potential with the Gaussian covariance function 
 given between (\ref{eq:Cpos}) and (\ref{eq:decayGauss}) in the limit $ \tau \downarrow 0 $. 
 In this limit (\ref{eq:decayGauss}) reduces to 
 $ \Gamma_\ll^2 = \left(\alpha^2 B /4 \pi \right) (2 \ll)!/(\ll ! \, 2^\ll)^2 $
 and the variance of $ \nu_\ll $ becomes, by (\ref{Eqbreite}), independent of 
 the Landau-level index, $ \sigma_\ll^2 = \sigma_0^2 = \alpha^2 B/(2\pi) $.
 Remarkably, in this limit explicit expressions for $ w_0 $ and $ w_\ll $, in the additional 
 high Landau-level limit $ \ll \to \infty $, are available.
 The first result is due to Wegner\cite{Weg83} and reads 
 \begin{equation}\label{Eq:Wegnerdelta}
       w_0\left( E \right) = 
       \frac{2}{\pi^{3/2} \sigma_0}  \,
       \frac{\exp(\eta^2)}{1 + \left[2 \pi^{-1/2} 
           \int_0^\eta \! {\rm d}\xi \, \exp(\xi^2) \right]^2}, 
       \qquad \eta := \frac{E}{ \sigma_0},
 \end{equation} 
 also see Refs.~\onlinecite{BeGrIt84,KlPe85,MaPu92}.
 Of course, when specializing the bound in (\ref{eq:Cpos}), it is
 consistent with (\ref{Eq:Wegnerdelta}) because $ 2 < \pi $.
 As for the second result, it is known\cite{BenCha86,Sal87} that $w_\ll$ 
 approaches for $ \ll \to \infty$ a semi-elliptic probability density, 
 \begin{equation}\label{eq:hll}
       \lim_{\ll \to \infty} w_\ll (E) = \frac{1}{2 \pi \sigma_0} \, 
       \Theta\left( 4 - \eta^2 \right) \, \sqrt{4  - \eta^2}, \qquad \eta = \frac{E}{ \sigma_0}.
 \end{equation}
 Unfortunately, in the delta-correlated limit the bound in (\ref{eq:Cpos}) 
 diverges asymptotically like $ \ll^{1/4}/(\pi^{1/4}\sigma_0) $ as $\ll \to \infty$.\\[1ex]
\indent
 (3)~  Different from (\ref{eq:hll}), for the above Gaussian covariance function
 with a strictly positive correlation length $ \tau > 0 $, the high Landau-level limit (informally)
 reads $ \lim_{\ll \to \infty} w_\ll (E) =  \delta(E) $. This follows from Chebyshev's inequality
 and the fact that the Landau-level broadening vanishes in this limit if $ \tau > 0 $:
 $ \lim_{\ll \to \infty} \, \sigma_\ll^2 = 0 $.\cite{BrHeLe91}
 In agreement with that, the bound in (\ref{eq:Cpos}) diverges in this case, 
 as may be seen either from $ 0 \leq \Gamma_\ll^2 \leq \sigma_\ll^2 $, valid\cite{BrHeLe91}
 for any covariance function, or directly from (\ref{eq:decayGauss}).\\[1ex]
\indent
 (4)~ The existence of a bounded $w_0$ in the 
    delta-correlated limit of a Gaussian random potential  
    stands in contrast to situations with random point impurities,
     $V^{(\omega)}(x) = \sum_j \lambda_j^{(\omega)} \, \delta\big(x - p_j^{(\omega)}\big)$. 
     To our knowledge, the following four cases have been considered so far:
     \begin{itemize}
     \item[(a)]
       the \emph{impurity positions}
       $p_j\in {{\mathbbm{R}}}^2$ randomly located according to Poisson's distribution
       and the \emph{coupling strengths} $\lambda_j \in {\mathbbm{R}} $  
       non-random, strictly positive, and all equal. \cite{BeGrIt84,Erd98}
     \item[(b)]
       $p_j \in {{\mathbbm{R}}}^2$ randomly located according to Poisson's distribution and 
       $\lambda_j  \in {\mathbbm{R}} $ independently, identically distri\-buted\cite{BeGrIt84} 
       according to a probability measure 
       whose support is a compact interval containing the origin.\cite{PuSc97}
     \item[(c)]
       $p_j \in {\mathbbm{Z}}^2$ non-random and 
       $\lambda_j  \in {\mathbbm{R}} $ independently, identically distributed according to a 
       bounded probability density 
       whose support is a compact interval containing the origin. \cite{DoMa97}
     \item[(d)]
       $p_j= j + d_j$ with $j \in {\mathbbm{Z}}^2$ non-random and the \emph{displacements} 
       $d_j \in{{\mathbbm{R}}}^2$ 
       independently, identically distributed  
       according to a bounded probability density with support contained in the 
       unit square $ ]-1/2,1/2[^2 \subset {\mathbbm{R}}^2 $. 
       Moreover, $\lambda_j  \in {\mathbbm{R}} $ as in the previous case. \cite{Scr99}
     \end{itemize} 
     In either of these cases, it has been shown
     that $ P_0 V^{(\omega)} P_0 $ has an infinitely degenerate eigenvalue at zero energy 
     for $ \mathbbm{P} $-almost all $ \omega \in \Omega $, if the magnetic-field strength 
     $B$ is sufficiently large.


%
\subsection{Gaussian upper bound on the restricted density of states}
As already pointed out, the estimate (\ref{ranzgauss}) 
is rather rough, because it does not depend on the energy.
Fortunately, under an additional isotropy assumption 
and with more effort one may construct an energy-dependent estimate.\\

\noindent
{\bf Theorem~2.}~ {\itshape
  Suppose 
  the situation of Theorem~1 and that the there defined convolution $ C_\mu $ 
  is spherically symmetric
  (with respect to the origin).
  Then the $\ll^{\rm th}$ restricted density of states $ w_\ll $ is bounded by a Gaussian 
  in the sense that
  \begin{equation}\label{Eqrestr}
    w_\ll(E) \leq \frac{1}{\sqrt{2\pi} \,
      \langle \psi_{\ll,0}\, , \,  C_\mu \, \psi_{\ll,0} \rangle} \,
      \exp\left(-\frac{E^2}{2C(0)}\right)
  \end{equation}
  for Lebesgue-almost all energies $ E \in {\mathbbm{R}} $. 
  {\rm [}Here $ \psi_{\ll, 0} $ is defined in (\ref{defpsi}).{\rm ]}
 }\\


\noindent
{\bf Remarks.}~ 
   (1)~ Equality holds in (\ref{Eqrestr}) (and (\ref{ranzgauss2}) below) with
   $ \langle \psi_{\ll,0}\, , \,  C_\mu \, \psi_{\ll,0} \rangle = \sqrt{C(0)} = \sigma_\ll = \Gamma_\ll $
   in the simple extreme case of a spatially constant Gaussian random potential $ V $, 
   that is, if $C(x) = C(0)$ for all $x \in {{\mathbbm{R}}}^2$. Of course, $ V $ is not 
   ergodic in this case. For a lucid discussion of ergodicity and related notions 
   in the theory of random (Schr{\"o}dinger) operators, see Ref.~\onlinecite{Kir89}.\\[1ex]
\indent
   (2)~ In view of (\ref{eq:tail}), we conjecture 
   the true leading decay of $w_\ll(E) = d \nu_\ll( ] - \infty , E [) / d E $ 
   for $|E| \to \infty$ to be
    Gaussian with decay energy $ \Gamma_\ll $. This energy 
    is strictly smaller than
    $ \sqrt{ C(0) }$, if not $C(x) = C(0)$ for all~$x \in {{\mathbbm{R}}}^2$.\\[1ex]
\indent
   (3)~ Using  
   $ \exp\big(-E^2/ 2C(0)\big) \leq 1 $ in (\ref{Eqrestr}), one obtains an energy-independent estimate
   which in general is weaker than (\ref{ranzgauss}) because
   $ \langle \psi_{\ll,0}\, , \, C_\mu \, \psi_{\ll,0} \rangle \leq \left\| P_\ll C_\mu P_\ll \right\| $.
   In particular this is true in the delta-correlated limit in which the energy-dependence of the 
   bound in~(\ref{Eqrestr}) disappears anyway.\\

\noindent
Gaussian random potentials with positive, spherically symmetric covariance functions illustrate
Theorem~2.\\

\noindent
{\bf Example~3.}~ 
   For a positive covariance function $ 0 \leq C(x) < \infty $, which is additionally 
   spherically symmetric,  
   the prefactor of the Gaussian in (\ref{Eqrestr}) is minimized by taking 
   $ \mu({\rm d}^2x) \, \gamma(\psi_{\ll,0}) = {\rm d}^2x \, \left| \psi_{\ll,0}(x) \right|^2 $ 
   so that $\langle \psi_{\ll,0}\, , \, C_\mu \, \psi_{\ll,0} \rangle = \gamma(\psi_{\ll,0}) $.
   By the Fourier representation~(\ref{eq:fourier}) and Jensen's inequality, with $ \widetilde C /C(0) $ 
   as the underlying Borel probability measure on $ {{\mathbbm{R}}}^2 $, one finds that
   $ \gamma^2(\psi_{\ll,0}) \geq \sigma_\ll^4/C(0) $. Therefore, the estimate~(\ref{Eqrestr}) 
   may be weakened to the following more explicit one
   \begin{equation}\label{ranzgauss2}
    w_\ll(E) \leq \frac{C(0)}{ \sigma_\ll^2}\,\frac{1}{\sqrt{2\pi C(0)}} \, 
    \exp\left(-\frac{E^2}{2C(0)}\right),
  \end{equation}
  where $ \sigma_\ll^2 $ is the variance of $ \nu_\ll $, see~(\ref{Eqbreite}).
  Alternatively, (\ref{ranzgauss2}) may be obtained directly from (\ref{Eqrestr}) by choosing
  $ \mu({\rm d}^2 x) \, \sqrt{C(0)} = {\rm d}^2 x \, \delta(x) $ so that $ C_\mu(x) = C(x)/\sqrt{C(0)} $.\\


\noindent
{\bf Remark.}~ 
    For the Gaussian covariance function $C(x) = C(0) \, \exp\big[ - \left| x \right|^2 / (2\tau^2) \big]$, 
    it is known\cite{Ape87,BrHeLe91} that $  \gamma^2(\psi_{0,0}) = \Gamma_0^2   
    = C(0) \, B \tau^2/(B \tau^2 +2) $, also see (\ref{eq:decayGauss}). 
    Theorem~2 together with the minimizing result mentioned in Example~3
    therefore gives the estimate
    \begin{equation}\label{eq:boehm}
       w_0(E) \leq \sqrt{\frac{B \tau^2 +2}{B \tau^2}} \, \frac{1}{\sqrt{2\pi C(0)}} \, 
       \exp\left(-\frac{E^2}{2C(0)}\right)
    \end{equation}
    for the restricted density of states of the lowest Landau band.
    In this setting $w_0$ has been approximately constructed using a continued-fraction 
    approach.\cite{BoBrLe97} 
    In accordance with the first remark below (\ref{Eqrestr}), a comparison with this approximation supports the fact
    that 
    the estimates~(\ref{eq:boehm}), (\ref{ranzgauss2}), and (\ref{Eqrestr})
    are the sharper, the longer the distance is over which the fluctuations of the Gaussian random potential
    are significantly correlated, more precisely, the larger the squared length ratio $ B \tau^2 $ is.\\
%
   
%
%
%
%

%
\section{Proof of the Gaussian upper bound}
\label{AppRest}
The proof of Theorem~2 requires two major ingredients, 
an approximation result (Proposition~2) and a  Wegner-type 
of estimate (Proposition~3).
We defer the details and proofs of these results to
Subsections~\ref{subsecApp}  
and \ref{subsecWeg}.
Taking these results for granted, the arguments for the validity of 
Theorem~2 are as follows.\\

\noindent
{\bf Proof of Theorem~2.}~
  Since the restricted density-of-states measure is even  
  for a (zero-mean) Gaussian random potential,
  that is, $ \nu_\ll(I)= \nu_\ll(-I)$ for all $I \in {\mathcal B}({{\mathbbm{R}}})$, and since we already know 
  from Theorem~1 that the density of states $ w_\ll $ exists and is bounded by a constant which does not exceed the 
  prefactor of the Gaussian in (\ref{Eqrestr}),
  it is sufficient to consider $ \nu_{\ll}$ on the strictly negative half-line $ ] - \infty , 0 [$.
  
  We now use Proposition~2 to show that a suitably defined 
  sequence of probability measures $ \left( \nu_{\ll,\nn} \right)_{n \in \mathbbm{N}}$ (see (\ref{nuN}) below) 
  converges weakly to $ \nu_{\ll}$ as $ n \to \infty $. 
  Given $E_1 < E_2 \leq 0$, we introduce the open interval $I:=]E_1, E_2[$. Then we have 
  \begin{equation}
     \nu_{\ll}(I) \leq  \liminf_{\nn \to \infty} \,  \nu_{\ll,\nn}(I), 
  \end{equation}
  by the portmanteau theorem (Thm. 30.10 in Ref.~\onlinecite{Bau92}).
  We now use Proposition~3 to estimate the prelimit expression and obtain 
  \begin{equation}
     \nu_{\ll}(I) 
    \leq   \frac{\left|I\right|}{\sqrt{2\pi} \,
         \langle \psi_{\ll,0} \, , \, C_\mu \, \psi_{\ll,0} \rangle} \,
     \exp\left(\beta E + \frac{\beta^2}{2}C(0) \right),
  \end{equation} 
  for all $ E \in [ E_2, 0] $ and all $\beta \geq 0$. Choosing $\beta = -
  E/C(0) \geq 0$ gives the claimed upper bound on $ w_\ll $ for $ E < 0 $. 
   \qed \\                               

\noindent
Before we proceed with the proofs of the approximation result and the  Wegner-type of estimate,
which were needed in the above proof,
we collect some preparations in the next subsection. 
                                %
\subsection{Angular-momentum eigenfunctions}\label{subsecAng}
The functions
\begin{equation}\label{EqAngular}
  x \mapsto \varphi_{\ll,k}(x):=  \sqrt{\frac{\ll!}{(\ll+k)!}} 
  \left[\sqrt{\frac{B}{2}}\, (x_1 +  i x_2)\right]^k 
  {\rm L}^{(k)}_\ll\left(\frac{B \,|x|^2}{2}\right) 
  \, \sqrt{\frac{B}{2\pi}} \,\exp\left(- \frac{B\, |x|^2}{4}\right)
\end{equation}
constitute\cite{Foc28} with $k \in \left\{-\ll,-\ll+1,\dots\right\}$ an
orthonormal basis in the $\ll^{\rm th}$ Landau-level 
eigenspace $P_\ll {\rm L}^2({\mathbbm{R}}^2)$.
In fact, $\varphi_{\ll,k}$ is an \emph{eigenfunction of the (perpendicular component 
of the canonical) angular-momentum} 
operator $L_3 := i \left(x_2 \partial/\partial x_1 - x_1 \partial/\partial x_2 \right)$
corresponding to the eigenvalue $k$, that is, $L_3 \varphi_{\ll,k} =  k  \, \varphi_{\ll,k}$.\\

\noindent
{\bf Lemma~2.}~ {\itshape
  Let $u:{\mathbbm{R}}^2 \to [ 0,\infty[$ be a measurable, positive, bounded, and spherically symmetric function. 
  Then the operator inequality 
  \begin{equation}
    P_\ll u_x  P_\ll 
    \geq \langle \psi_{\ll,0} \, , \, u \, \psi_{\ll,0} \rangle  \; \psi_{\ll,x} \langle \psi_{\ll,x} \, , \cdot \, \rangle
  \end{equation}
  holds for all $x \in {{\mathbbm{R}}}^2$.
  Here the function $ u_x (\, \cdot \,) := u(\, \cdot \, - x) $ is the $ x $-translate 
  of $ u $ and $  \psi_{\ll,x} \langle \psi_{\ll,x} \, , \cdot \, \rangle $ denotes 
  the orthogonal projection operator onto 
  the one-dimensional subspace spanned by $ \psi_{\ll,x} $, see~(\ref{defpsi}).  
}\\

\noindent
{\bf Proof.}~ 
  Since the function $u$ is spherically symmetric, 
  the operator $P_\ll u P_\ll$ is diagonal in the angular-momentum basis such that
  \begin{equation}\label{eqdiag}
    P_\ll u P_\ll = \sum_{k=-\ll}^\infty \langle \varphi_{\ll,k}, u \, \varphi_{\ll,k} \rangle \; 
    \varphi_{\ll,k}  \langle \varphi_{\ll,k}\, , \cdot \, \rangle 
    \geq 
    \langle \varphi_{\ll,0}, u \, \varphi_{\ll,0} \rangle \,  \varphi_{\ll,0}  \langle \varphi_{\ll,0}\, , \cdot \, \rangle.
  \end{equation}
  The shifted operator $ P_\ll u_x P_\ll =  T_x \, P_\ll \, u \, P_\ll\, T_x^\dagger $ 
  results from the l.h.s.\ of (\ref{eqdiag}) by a unitary
  transformation with the magnetic translation $T_x$, see 
  (\ref{eq:magntrans}). The proof is hence completed by  
  observing that $\varphi_{\ll,0}= \psi_{\ll,0}$ and $\psi_{\ll,x} = T_x \psi_{\ll,0}$. \qed \\
                                %

Subsequently, we will consider the 
$\nn$-dimensional subspaces
$
P_{\ll,\nn}{\rm L}^2({\mathbbm{R}}^2)\subset P_\ll{\rm L}^2({\mathbbm{R}}^2)
$
spanned by the first $\nn$ angular-momentum eigenfunctions. The orthogonal
projection $P_{\ll,\nn}$ is therefore defined by
\begin{equation}
  P_{\ll,\nn} := \sum_{k=-\ll}^{\nn-\ll-1}  \varphi_{\ll,k}  \langle \varphi_{\ll,k}\, , \cdot \, \rangle,
  \qquad \nn \in \mathbbm{N}. 
\end{equation}
The completeness of $ \{ \varphi_{\ll,k} \} $ in $  P_\ll{\rm L}^2({\mathbbm{R}}^2)$ implies
the strong-limit relation $\textrm{s-}\lim_{\nn\to\infty}P_{\ll,\nn}=P_{\ll}$ on $ {\rm L}^2({\mathbbm{R}}^2) $. 
The projections $P_{\ll,\nn}$ 
are integral operators with (continuous) kernels $P_{\ll,\nn}(x,y)$, whose diagonals are 
given by $ P_{\ll,\nn}(x,x)= B\,  G_{\ll,\nn}\!\left(B\,|x|^2/2\right) /2 \pi \leq B/ 2 \pi $. 
Here the function
\begin{equation}\label{DefG}
  G_{\ll,\nn}\left( \xi \right) :=  {e}^{- \xi} \, \sum_{k= - \ll}^{\nn- \ll -1} \frac{ \ll !}{(k+ \ll)!} \, \xi^k \, 
  \left( {\rm L}^{(k)}_\ll(\xi) \right)^2, \qquad \xi \geq 0, \qquad \nn \in \mathbbm{N},
\end{equation}
is approximately one and approximately zero for $ \xi $ smaller and 
larger than $ \nn - 1/2 $, respectively. 
Moreover, the length of the interval on which its values 
differ significantly from one and zero does not depend on $\nn$, also see the remark after
the following\\

\noindent                              %
{\bf Lemma~3.}~ {\itshape
  Let $G_{\ll,\nn}$ be defined by (\ref{DefG}). 
  Then the following scaling-limit relation holds 
  \begin{equation}\label{Eq:Scal}
    \lim_{\nn \to \infty} G_{\ll,\nn}(\nn \xi) = \left\{ \begin{array}{l@{$\qquad$\mbox{\rm if}$\qquad$}l}
        1 & 0 < \xi < 1 \\
        0 & 1 < \xi < \infty.
      \end{array}\right.        
  \end{equation}
  Moreover, for every $ \ll \in {\mathbbm{N}}_0 $ there exist an $N_\ll \in {\mathbbm{N}} $ and a real $A_\ll > 0$, 
  such that
  $ 0 \leq  G_{\ll,\nn}(\nn \xi) \leq A_\ll \, {e}^{- \xi} $ for all $ \xi \geq 0 $ and 
  all $ \nn \geq N_\ll $.}\\

\noindent
{\bf Remark.}~ 
  With more effort one may even prove that for every $ \ll \in {\mathbbm{N}}_0 $
  there exists some polynomial $ \zeta \mapsto {\rm Pol}(\zeta,\ll ) $ of maximal degree $ 2 \ll + 1 $ such that
  \begin{equation}
     0  \leq  G_{\ll,\nn}\!\left( ( \sqrt{\nn - 1/2} + \zeta)^2 \right)  
     \leq  {e}^{- \zeta^2} \, {\rm Pol}(\zeta,\ll )   
  \end{equation}
  for all $ \nn \in \mathbbm{N} $ and all $ \zeta \geq 0 $. Moreover,
  \begin{equation}
     1 -  {e}^{- \zeta^2} \,  {\rm Pol}(-\zeta,\ll ) \leq  G_{\ll,\nn}\!\left( ( \sqrt{\nn - 1/2} - \zeta)^2 \right) 
     \leq 1 
  \end{equation}
  for all $ n \in {\mathbbm{N}} + \ll $ and all $ 0 \leq \zeta \leq \sqrt{\nn - 1/2 } $.\\

\noindent
{\bf Proof of Lemma~3.}~ 
  The proof is based on the following recurrence relation
  \begin{equation}\label{Eq:Rec}
     G_{\ll,\nn}(\xi) -  G_{\ll -1,\nn}(\xi) =
     - \, {e}^{- \xi} \, \frac{(\ll -1)!}{(\nn -1)!} \, \xi^{\nn - \ll}  \,
     {\rm L}^{(\nn - \ll)}_{\ll-1}(\xi) \, {\rm L}^{(\nn - \ll)}_{\ll}(\xi) 
     =:  D_{\ll,\nn}(\xi)
  \end{equation}
  for all $ \ll \geq 1 $. 
  It follows from the fact that $ D_{\ll,\nn} $ may be written as a 
  telescope sum according to
  \begin{equation}\label{Eq:Tele}
   D_{\ll,\nn}(\xi) =  {e}^{- \xi} \sum_{k = - \ll + 1}^{\nn - \ll} 
   \frac{ ( \ll -1)!}{(k + \ll - 1)!} \, \xi^k \, 
   \Big[ \, \frac{ k + \ll -1}{\xi} \, 
     {\rm L}^{(k - 1)}_{\ll-1}(\xi) \, {\rm L}^{(k - 1)}_{\ll}(\xi) 
     -  {\rm L}^{(k)}_{\ll-1}(\xi) \, {\rm L}^{(k)}_{\ll}(\xi) \Big].
   \end{equation}
   Equation~8.971(4) in Ref.~\onlinecite{GrRy} may be written as 
   $ (k + \ll - 1 ) \, {\rm L}^{(k - 1)}_{\ll-1}(\xi) 
   = \xi \, {\rm L}^{(k)}_{\ll-1}(\xi) + \ll \, {\rm L}^{(k-1)}_{\ll}(\xi) $.
   Using this together with Eq.~8.971(5) in Ref.~\onlinecite{GrRy}, 
   the difference in the square bracket in (\ref{Eq:Tele})
   is seen to be equal to 
   $ \big( {\rm L}^{(k - 1)}_{\ll}(\xi) \big)^2 \, \ll/\xi - \big( {\rm L}^{(k)}_{\ll-1}(\xi) \big)^2 $.
   Splitting the sum into two parts yields (\ref{Eq:Rec}).
   The proof of (\ref{Eq:Scal}) then
   follows by mathematical induction on $ \ll \in {\mathbbm{N}}_0 $. 
   In case $ \ll = 0 $ we write
   \begin{equation}\label{Eq:null}
      G_{0,\nn}(\xi) =  {e}^{- \xi} \sum_{k=0}^{\nn -1} \frac{\xi^k}{k!} 
      =: {e}^{- \xi} \, {e}_\nn(\xi) - {e}^{- \xi} \, \frac{\xi^\nn}{\nn!}
   \end{equation}
   in terms of the incomplete exponential function $ {e}_\nn $, 
   (Eq.~6.5.11 in Ref.~\onlinecite{AbrSteg}).
   By Stirling's estimate $ \nn! \geq \sqrt{ 2 \pi \nn} \, \nn^\nn {e}^{-\nn} $ 
   for the factorial
   (Eq.~6.1.38 in Ref.~\onlinecite{AbrSteg}) 
   and the elementary inequality
   $ \xi - 1 - \ln \xi \geq 0$,
   the second term on the r.h.s. of (\ref{Eq:null}) 
   vanishes in the scaling limit (\ref{Eq:Scal}) 
   such that the claim reduces 
   to the content of Eq.~6.5.34 in  Ref.~\onlinecite{AbrSteg} for $ \ll = 0 $.
   For the induction clause, we use the following
   exponential, hence rough, growth limitation for Laguerre polynomials
   \begin{equation}
     \big|  {\rm L}^{(k)}_{\ll}(\xi) \big| 
     = \big| \, \sum_{j=0}^\ll (- 1)^j \, \binom{\ll + k}{\ll -j} \, \frac{\xi^j}{j!} \, \big|
     \leq 
     \sum_{j=0}^\ll ( \ll + k)^{\ll - j} \, \frac{\, \xi^j}{j!}
                                           \leq ( \ll + k )^\ll \, {e}^{\xi/(\ll + k)}
   \end{equation}
   which is valid for $k \geq 1 - \ll $ and obtained by bounding the binomial coefficients. 
   Using again Stirling's estimate, this yields the inequality
   \begin{equation}\label{Eq:SchrankeD}
     \left|   D_{\ll,\nn}(\nn \xi) \right| \leq ( \ll -1 )! \, {e}^{3 \xi} \, \left( \frac{\nn}{\xi} \right)^\ll \,
     {e}^{ - (\xi - 1 - \ln \xi )\, n }
   \end{equation}
   for all $ \ll \geq 1 $ and all $ \nn \geq 2 $. Since $ \xi - 1 - \ln \xi > 0 $ for all $ \xi \neq 1 $,
   we have $ \lim_{n \to \infty} D_{\ll,\nn}(\nn \xi) = 0 $ and hence
   $  \lim_{\nn \to \infty}  G_{\ll,\nn}(\nn \xi) =  \lim_{\nn \to \infty}  G_{\ll-1,\nn}(\nn \xi) $
   for all $  \xi \neq 1 $, which completes the proof of  (\ref{Eq:Scal}).
   
   For a proof of the
   exponential bound $ 0 \leq   G_{\ll,\nn}(\nn \xi) \leq A_\ll {e}^{- \xi} $ 
   with some $ A_\ll > 0 $ and
   $ \nn $ large enough, we first recall that
   \begin{equation}
     0 \leq  G_{\ll,\nn}(\xi) \leq  G_{\ll,\infty}(\xi) = 1
   \end{equation}
   for all $ \xi \geq 0 $, $ \ll \in {\mathbbm{N}}_0 $, and $ \nn \in {\mathbbm{N}} $.
   Using $\nn^k \leq (\nn-1)^k\,{e}$ for $0\leq k\leq n-1 $ in (\ref{Eq:null}),
   one obtains $G_{0,\nn}(\nn \xi) \leq   {e}^{1 - \xi} $ for all $ \xi \geq 0 $.
   The claimed exponential bound
   for all $\ll \in {\mathbbm{N}}_0$ then follows from 
   (\ref{Eq:SchrankeD}) and (\ref{Eq:Rec}). 
   \qed
  %
                                %
\subsection{Approximating sequence of probability measures on the real line}\label{subsecApp}
Employing the  $\nn\times \nn$ random Hermitian matrices 
$P_{\ll,\nn} V^{(\omega)} P_{\ll,\nn}$,
we define a sequence $ (\nu_{\ll,\nn})_{n \in {\mathbbm{N}}} $ of probability measures by 
\begin{equation}\label{nuN}
   \nu_{\ll,\nn}(I) :=  
  \frac{1}{\nn}\,{\mathbbm{E}}\left\{ \Tr\left[P_{\ll,\nn} \, 
      \indfkt{I}(P_{\ll,\nn} V P_{\ll,\nn}) \, P_{\ll,\nn}\right]\right\},
  \quad I \in \mathcal{B}({\mathbbm{R}}).
\end{equation}
Here the trace $ \Tr\left[P_{\ll,\nn} \, 
      \indfkt{I}(P_{\ll,\nn} V^{(\omega)} P_{\ll,\nn}) \, P_{\ll,\nn}\right] $ is equal
to the (random) number of eigenvalues
(counting multiplicity) of $ P_{\ll,\nn} V^{(\omega)} P_{\ll,\nn} $ in the Borel set $ I $.
For rather general random potentials the sequence $ (\nu_{\ll,\nn}) $ approximates the
restricted density-of-states measure $ \nu_\ll$.
This is the first ingredient of the proof of Theorem~2.\\

\noindent
{\bf Proposition 2.}~ {\itshape
  Let $V$ be an ${\mathbbm{R}}^2$-homogeneous random potential with ${\mathbbm{E}}\{|V(0)|\} < \infty$. 
  Moreover, assume that
  $P_{\ll} V^{(\omega)} P_{\ll}$ and $ P_{\ll,\nn} V^{(\omega)} P_{\ll,\nn} $ for all $ \nn \in {\mathbbm{N}} $
  are self-adjoint on $ {\rm L}^2({\mathbb{R}}^2) $ 
  for $\mathbbm{P}$-almost all $\omega \in \Omega $. Then 
  \begin{equation}
     \nu_\ll = \lim_{\nn\to \infty} \,  \nu_{\ll,\nn}
  \end{equation}
  in the sense of weak convergence of finite measures.
}\\

\noindent                                %
{\bf Remark.}~ 
  The assumptions of the proposition are fulfilled for a Gaussian random potential
  in the sense of Definition~1, because $ P_{\ll} V P_{\ll} $ is almost surely 
  essentially self-adjoint on $ {\mathcal{S}}({\mathbbm{R}}^2) $ by Proposition~1. 
  Moreover, the random matrix operator $ P_{\ll,\nn} V P_{\ll,\nn} $ 
  is almost surely self-adjoint for all $ \nn \in {\mathbbm{N}} $, because of the 
  almost-sure finiteness
  $\left| \langle \varphi_{\ll,j} , V \varphi_{\ll,k} \rangle \right| < \infty $ 
  for all $ j, k \in {\mathbbm{N}}_0 - \ll $.\\

\noindent
{\bf Proof of Proposition 2.}~ 
  The claimed weak convergence of (finite) measures
  is equivalent to pointwise convergence of their Stieltjes transforms, that is,
  \begin{equation}\label{Stieltjes}
    \lim_{\nn\to \infty} \int_{{\mathbbm{R}}}\! 
    \frac{\nu_{\ll,\nn}({\rm d}E)}{E - z} = 
    \int_{{\mathbbm{R}}}\! \frac{\nu_{\ll}({\rm d}E)}{E - z}
  \end{equation}
  for all $z\in\mathbbm{C}\backslash{\mathbbm{R}}$, 
  see, for example, Prop.~4.9 in Ref.~\onlinecite{PaFi92}. 
  The spectral theorem and (\ref{Schur}) show that the latter convergence
  follows from
  \begin{equation} \label{Stieltjes2}
    \lim_{\nn\to\infty} \, \frac{1}{\nn} \, {\mathbbm{E}} \left\{
      \left|\Tr\left[P_{\ll,\nn} 
          \left((P_{\ll,\nn} V P_{\ll,\nn}-z)^{-1}
            - (P_{\ll} V P_{\ll} -z)^{-1} \right)  
          P_{\ll,\nn}\right]\right| 
    \right\} 
    = 0. 
  \end{equation}
  As a self-adjoint operator of finite rank, $ P_{\ll,\nn} V^{(\omega)} P_{\ll,\nn} $ is 
  defined on the whole space $ {\rm L}^2({\mathbbm{R}}^2) $ for $ \mathbbm{P} $-almost all $ \omega \in \Omega $,
  so that we may use the (second) resolvent equation.\cite{Weid80} 
  Together with the fact that $(P_{\ll,\nn} V^{(\omega)} P_{\ll,\nn} - z )^{-1}$ 
  and $P_{\ll,\nn}$ commute with each other, the absolute value of the trace in (\ref{Stieltjes2}) is hence 
  seen to be equal to
  \begin{align}\label{Holder}
   & \left|\Tr\left[P_{\ll,\nn} (P_{\ll,\nn} V^{(\omega)} P_{\ll,\nn} - z)^{-1}
          P_{\ll,\nn} V^{(\omega)} (P_{\ll} - P_{\ll,\nn}) 
          (P_{\ll} V^{(\omega)} P_{\ll} - z )^{-1} P_{\ll,\nn}\right]\right| 
      \notag \\
    & \qquad  \leq \left\|(P_{\ll,\nn} V^{(\omega)} P_{\ll,\nn} - z)^{-1}\right\|\, 
    \left\|( P_{\ll} V^{(\omega)} P_{\ll} -z)^{-1}\right\| \, 
    \left\|P_{\ll,\nn} V^{(\omega)} (P_{\ll} - P_{\ll,\nn})\right\|_1 
      \notag \\
    & \qquad  
    \leq |\Im z|^{-2} \left\|P_{\ll,\nn} V^{(\omega)} (P_{\ll} - P_{\ll,\nn})\right\|_1. 
  \end{align}
  Here we employed H{\"o}lder's inequality for the 
  trace norm $ \| A \|_1:= \Tr \left( A^{\dagger} A  \right)^{1/2} $ and  
  $\Im z$ denotes the imaginary part of $z$. 
  The trace norm in (\ref{Holder})
  is in turn estimated as follows
  \begin{align}
    \left\|P_{\ll,\nn} V^{(\omega)} (P_{\ll} - P_{\ll,\nn})\right\|_1 
    & = \Big\| \, \frac{B}{2\pi}\,\int_{{{\mathbbm{R}}}^2} \!\! {\rm d}^2{x}\, \,  V^{(\omega)}(x)\, 
    \psi_{\ll,x,\nn} \, \langle \psi_{\ll,x} - \psi_{\ll,x,\nn}, \cdot \, \rangle\,  \Big\|_1 
    \notag \\
    & \leq \frac{B}{2\pi}\,\int_{{{\mathbbm{R}}}^2} \!\! {\rm d}^2{x} \left|V^{(\omega)}(x)\right|  
    \big\|\, \psi_{\ll,x,\nn} \, \langle \psi_{\ll,x} - \psi_{\ll,x,\nn}, \cdot \, \rangle \, \big\|_1 
    \notag \\
    & =  \int_{{{\mathbbm{R}}}^2} \!\! {\rm d}^2{x} \left|V^{(\omega)}(x)\right| \sqrt{P_{\ll,\nn}(x,x) \left[P_{\ll}(x,x) - P_{\ll,\nn}(x,x)\right]},
    \label{Tracenorm}
  \end{align}
  where we introduced the sequence of two-parameter families of complex-valued functions
  \begin{equation}\label{Def:psin}
    y \mapsto \psi_{\ll,x,\nn}(y):= \left(P_{\ll,\nn}\, \psi_{\ll,x}\right)(y), \qquad x \in {\mathbbm{R}}^2.
  \end{equation}
  Note that these functions are not normalized,  
  $\langle \psi_{\ll,x,\nn}\, , \psi_{\ll,x,\nn} \rangle = 2\pi P_{\ll,\nn}(x,x)/B = G_{\ll , \nn}(B | x |^2 / 2 ) \leq 1$.
  Combining (\ref{Holder}) and (\ref{Tracenorm}), using Fubini's theorem and the
  homogeneity of the random potential, the l.h.s.\ of 
  (\ref{Stieltjes2}) is seen to be bounded from above by
  \begin{multline}
    \lim_{\nn\to\infty} \; \frac{{\mathbbm{E}}\left[| V(0) | \right]}{\nn \, \left|\Im z \right|^2} 
    \int_{{{\mathbbm{R}}}^2} \!\! {\rm d}^2{x} \sqrt{P_{\ll,\nn}(x,x) \left[P_{\ll}(x,x) - P_{\ll,\nn}(x,x)\right]} \\
    = \frac{{\mathbbm{E}}\left[| V(0) | \right]}{\left|\Im z \right|^2} \;
     \lim_{\nn\to\infty} \; \int_0^\infty\!{\rm d}\xi \, \sqrt{G_{\ll,\nn}(\nn \xi) \left[1 - G_{\ll,\nn}(\nn\xi)\right]}
     = 0. \qquad
  \end{multline} 
  Here we employed the definition of $G_{\ll,\nn}$ (see text above (\ref{DefG})), performed the angular integration,
  changed variables $\nn\xi :=B |x|^2/2$ in the remaining integral, and used 
  Lemma~3 and the dominated-convergence theorem. \qed
                                %
  %
                                %
\subsection{A Wegner-type of estimate}\label{subsecWeg}
The second ingredient for the proof of Theorem~2 is the following\\
 
\noindent
{\bf Proposition 3.}~ {\itshape
  In the situation of Theorem~2 let $E_1 < E_2 \leq E \leq 0$ and put $I:=]E_1, E_2[$. Then
  \begin{equation}\label{eqRWeg}
     \nu_{\ll,\nn}(I) \leq \left(  \frac{\left|I\right|}{\sqrt{2\pi} \,
         \langle \psi_{\ll,0} \, , \, C_\mu \, \psi_{\ll,0} \rangle}  + s_{\ll,\nn} \right)\,
     \exp\left(\beta E + \frac{\beta^2}{2}C(0) \right),
  \end{equation}
  for all $\beta \geq 0$.
  Here $s_{\ll,\nn} := \int_1^\infty \! {\rm d} \xi \, G_{\ll,\nn}(\nn\xi) $
  converges to zero as $\nn\to \infty$. } \\

\noindent
{\bf Proof.}~ 
  By the 
  definition of $ \nu_{\ll,\nn}$ and the spectral theorem one has
  \begin{equation}
     \nu_{\ll,\nn}(I) 
    \leq
    \frac{{e}^{\beta E}}{\nn}  \, {\mathbbm{E}}\left\{ \Tr\left[P_{\ll,\nn} \, 
      {e}^{- \beta P_{\ll,\nn} V P_{\ll,\nn}} 
      \indfkt{I}(P_{\ll,\nn} V P_{\ll,\nn}) \, P_{\ll,\nn}\right] \right\}.
    \label{eqeinschub}
  \end{equation}
  We evaluate the trace in an orthonormal eigenbasis 
  of $ P_{\ll,\nn} V^{(\omega)} P_{\ll,\nn} $ and apply the 
  Jensen-Peierls inequality\cite{Ber72} to bound the probabilistic expectation in (\ref{eqeinschub}) 
  from above by
  \begin{multline}
    {\mathbbm{E}}\left\{ \Tr\left[P_{\ll,\nn} \, 
      {e}^{- \beta V } P_{\ll,\nn} \, 
      \indfkt{I}(P_{\ll,\nn} V P_{\ll,\nn}) \, P_{\ll,\nn} \right] \right\} \\
    = \frac{B}{2\pi} \, \int_{{{\mathbbm{R}}}^2} \!\! {\rm d}^2{x} \,\, 
    {\mathbbm{E}}\left\{ \, {e}^{- \beta V(0) } \, 
    \langle\psi_{\ll,x,\nn}\, , \, \indfkt{I}(P_{\ll,\nn} V( \, \cdot \, - x ) P_{\ll,\nn}) \, 
    \psi_{\ll,x,\nn}\rangle  \right\} \qquad
    \label{eqpulldown}
  \end{multline}
  where we used Fubini's theorem and the $ {\mathbbm{R}}^2 $-homogeneity of $ V $.  
  The Lebesgue integral in (\ref{eqpulldown}) over the plane may be split into two parts with
  domains of integration inside and outside a disk centered at the origin and
  with radius $\sqrt{2\nn/B}$. 

  To estimate the inner part, we use the 
  one-parameter decomposition (\ref{eq:decomposition}) of the
  Gaussian random potential $ V $. 
  Since $ U $ and $ \lambda $  
  are stochastically independent,
  we may estimate 
  the conditional expectation of the integrand in (\ref{eqpulldown}) 
  relative to the sub-sigma-algebra generated by
  $\{U(y)\}_{y\in {\mathbbm{R}}^2}$
  with the help of Lemma~1. Taking there 
  $g(\xi) = \exp\big(- \beta \xi \, C_\mu(0) - \xi^2 /2\big) / \sqrt{2\pi} $, 
  $K = \psi_{\ll,x,\nn} \, \langle \psi_{\ll,x,\nn}\, , 
  \, \cdot \rangle / \langle\psi_{\ll,x,\nn}, \, \psi_{\ll,x,\nn}\rangle$, and 
  $ M =  P_{\ll,\nn} C_\mu(\, \cdot \, - x) P_{\ll,\nn} 
  \geq \langle \psi_{\ll,0} \, , \, C_\mu \, \psi_{\ll,0} \rangle\, \langle\psi_{\ll,x,\nn}, \, \psi_{\ll,x,\nn}\rangle \, K^2 $, 
  where the last inequality
  follows from Lemma~2 
  and the  positivity as well as the  spherical symmetry of $ C_\mu $, we
  obtain an $ \omega $- and $ x $-independent bound according to
  \begin{multline}\label{eqCondExp}
     \mkern-10mu \int_{{\mathbbm{R}}}\! {\rm d} \xi \,\, 
     \frac{{e}^{-\xi^2/2}}{\sqrt{2\pi}} \, {e}^{- \beta \xi \, C_\mu(0)} \,
    \big\langle\psi_{\ll,x,\nn}\, , \, 
    \indfkt{I}\left(P_{\ll,\nn}U^{(\omega)}( \, \cdot \, - x )  P_{\ll,\nn} 
      + \xi\, P_{\ll,\nn} C_\mu(\, \cdot \, - x) P_{\ll,\nn} \right) \, 
    \psi_{\ll,x,\nn}\big\rangle \qquad \\
    \leq \frac{|I|}{\sqrt{2\pi} \, \langle \psi_{\ll,0} \, , \, C_\mu \, \psi_{\ll,0} \rangle}
    \, \exp\left(\frac{\beta^2}{2} \left(C_\mu(0)\right)^2\right).
    \qquad
  \end{multline} 
  Using $ {\mathbbm{E}}\big[ \exp\left( - \beta U(0) \right) \big] 
      = \exp\big( \beta^2 \big(C(0) - \left(C_\mu(0)\right)^2 \big)/2 \big) $,
  the inner  part of the integral in (\ref{eqpulldown}),
  may hence be estimated as follows
  \begin{multline}
    \int_{|x|^2 \leq 2\nn/B} \mkern-30mu {\rm d}^2 x \; 
     {\mathbbm{E}}\, \Big\{ \, {e}^{- \beta V(0) } \, 
      \langle\psi_{\ll,x,\nn}\, , \, 
      \indfkt{I}(P_{\ll,\nn} V(\, \cdot - x ) P_{\ll,\nn}) \, 
      \psi_{\ll,x,\nn}\rangle \Big\} \\
    \leq
     \frac{2 \pi \nn}{B} \, \frac{\left|I\right| }{\sqrt{2\pi} \,
       \langle \psi_{\ll,0} \, , \, C_\mu \, \psi_{\ll,0} \rangle} \,
    \exp\left(\frac{\beta^2}{2} C(0)  \right).
    \qquad
  \end{multline}
  This gives the first part of the claimed inequality (\ref{eqRWeg}). 

  To complete the proof, we estimate the outer part of the integral in (\ref{eqpulldown}) 
  as follows
  \begin{multline}
    \int_{|x|^2 \geq 2\nn/B} \mkern-30mu 
    {\rm d}^2 x \;  {\mathbbm{E}}\Big\{ \, {e}^{- \beta V(0) } \, 
      \langle\psi_{\ll,x,\nn}\, ,  
      \indfkt{I}(P_{\ll,\nn} V(\, \cdot - x ) P_{\ll,\nn}) \, \psi_{\ll,x,\nn}\rangle \Big\} \\
    \leq 
     {\mathbbm{E}}\left\{ {e}^{- \beta V(0) }\right\} \;
     \int_{|x|^2 \geq 2\nn/B} \mkern-30mu 
    {\rm d}^2 x \;\;  \langle\psi_{\ll,x,\nn}\, , \, \psi_{\ll,x,\nn}\rangle 
    =  \frac{2\pi \nn}{B} \, {\mathbbm{E}}\left\{ {e}^{- \beta V(0) }\right\} \; 
    \int_1^\infty \! {\rm d} \xi \, G_{\ll,\nn}(\nn\xi). 
    \label{eqremaining}
  \end{multline}
  Here we employed (\ref{Def:psin}) and changed variables $\nn \xi := B \, |x|^2/2$ to
  obtain the last equality.
  Thanks to Lemma~3, the last integral in (\ref{eqremaining}), and hence $ s_{\ll, \nn} $, 
  converges to zero
  as $\nn \to \infty$ by the dominated-convergence theorem. \qed
%

%

%
\section*{Acknowledgments}
Our thanks go to Peter M{\"u}ller, Olaf Rogalsky and Michael Sch{\"u}tz 
for helpful remarks. 
We thank the referee for comments that improved the presentation of the paper. 
This work was supported by the Deutsche Forschungsgemeinschaft under
grant no.~Le 330/12 within the
Schwerpunktprogramm ,,Interagierende stochastische Systeme von hoher
Komplexit{\"a}t`` (DFG Priority Program SPP 1033). 
%
%


%
%

%
%

\end{document}